\documentclass[preprint,aps,showpacs]{revtex4}
\usepackage{graphicx}
\begin{document}

\title{Stability of a two-sublattice spin-glass model}

\author{Carlos S. O. Yokoi}

\email{cyokoi@if.usp.br}

\affiliation{Instituto de F\'{\i}sica, Universidade de S\~{a}o Paulo,
  Caixa Postal 66318, 05315-970 S\~{a}o Paulo, SP, Brazil}

\author{Francisco A. da Costa}

\email{fcosta@dfte.ufrn.br}

\affiliation{Departamento de F\'{\i}sica Te\'orica e Experimental,
  Universidade Federal do Rio Grande do Norte, Caixa Postal 1641,
  59072-970 Natal, RN, Brazil}

\date{\today}

\begin{abstract}
We study the stability of the replica-symmetric solution of a 
two-sublattice infinite-range spin-glass model, which can describe the transition  
from antiferromagnetic to spin glass state.  The eigenvalues associated with 
replica-symmetric perturbations  are in general complex.  The natural generalization 
of the usual  stability condition is to require the real part of these eigenvalues  to 
be positive.  The necessary and sufficient conditions for all the  roots of the 
secular equation to have positive real parts is given  by the Hurwitz criterion.  The 
generalized stability condition  allows a consistent analysis of the phase diagram 
within the  replica-symmetric approximation.


\end{abstract}

\maketitle

\section{Introduction}

The infinite-range Sherrington-Kirkpatrick (SK) model
\cite{sherrington75} for a spin glass has attracted considerable
attention over the past decades \cite{binder86,mezard87,fischer91}.
These investigations have revealed highly non-trivial properties such
as the instability of replica-symmetric (RS) solution \cite{almeida78}
and the replica-symmetry-breaking scheme to produce a stable solution
\cite{parisi79,parisi80a,parisi80b,parisi80c}. Most studies have
concentrated on situations where the exchange distributions are either
symmetric or with an additional ferromagnetic interaction. More
recently a two-sublattice version of the SK model was introduced
\cite{korenblit85,fyodorov87a,fyodorov87b,takayama88} to allow for
antiferromagnetic interactions between different sublattices.  Such
extension is quite natural in view of the existence of many
experimental systems such as
Fe${}_x$Mg${}_{1-x}$Cl${}_2$\cite{bertrand82,wong85a,wong85b} and
Fe${}_x$Mn${}_{1-x}$TiO${}_3$\cite{yoshizawa87,yoshizawa89}, which
exhibit transition from and Ising antiferromagnetic into an Ising spin
glass state for certain range of $x$ values.  In contrast to the
standard SK model, in the two-sublattice SK model with
antiferromagnetic intersublattice interactions, the ordered
(antiferromagnetic) phase extends to finite fields and the de
Almeida-Thouless instability line \cite{almeida78} has distinct
branches in the paramagnetic and antiferromagnetic phases, which do not
meet at a first-order transition
\cite{korenblit85,fyodorov87a,fyodorov87b,takayama88}.  Experimental
determination of the field-temperature phase diagram in
Fe${}_x$Mn${}_{1-x}$TiO${}_3$, as well as the de Almeida-Thouless
instability line \cite{yoshizawa94}, are in qualitative agreement with
mean-field results \cite{takayama88}.

In the previous studies of this model the stability of the RS solution
against transversal fluctuations, i.e., outside the RS space, has
already been investigated
\cite{korenblit85,fyodorov87a,fyodorov87b,takayama88}, and the
stability against longitudinal fluctuations, i.e., inside the RS
space, was also briefly considered \cite{fyodorov87b}. The stability
of the RS solution against transversal fluctuations is important to
establish whether replica symmetry breaking is necessary. The
stability against longitudinal fluctuations, however, is also
necessary to ensure the validity of RS solution.  For certain
parameter values of the two-sublattice SK model there may be up to
three RS solutions, all of them stable against transversal
fluctuations.  In such a situation the analysis of the stability
against longitudinal fluctuations is important for a consistent study
of the phase diagram by eliminating unstable solutions.

In this work we remedy the lack of such investigation by a detailed
numerical and analytical study of the eigenvalues associated with
longitudinal fluctuations. Surprisingly, these eigenvalues are in
general complex.  It is natural to assume that stability condition
should require the real part of these eigenvalues to be positive.  The
necessary and sufficient condition for all the roots of the secular
equation to have positive real part is given by the Hurwitz criterion.
We show that this generalized stability condition allows a consistent
study of the phase diagram within the RS approximation.


\section{The model}

We consider a system of $2N$ Ising spins $S_i=\pm 1$ located at the
sites of two identical sublattices $A$ and $B$. The interactions are
described by the Hamiltonian
\begin{equation}
  {\cal H}=-\sum_{i \in A, j \in B} J_{ij} S_i S_j- \sum_{(ij) \in A}
  J^{\prime }_{ij} S_i S_j - \sum_{(ij) \in B} J^{\prime }_{ij} S_i S_j
  - H \sum_i S_i,
  \label{hamiltonian}
\end{equation}
where the first sum is over all distinct pairs of spins belonging to
different sublattices, the second and third ones refer to all distinct
pairs of spins belonging to the same sublattices, and the last sum is
over all spins in the two sublattices.  $J_{ij}$ is the exchange
interaction between spins in different sublattices, $J^{\prime }_{ij}$
is the exchange interaction between spins in the same sublattice, and
$H$ is the applied magnetic field.  The exchange interactions are
independent, quenched, Gaussian random variables with mean values
\begin{equation}
  \langle J_{ij} \rangle_J = \frac{J_0}{N},  \qquad \langle
  J^{\prime}_{ij} \rangle_J = \frac{J^{\prime}_0}{N},
\end{equation}
and variances
\begin{equation}
  \langle J_{ij}^2 \rangle_J - \langle J_{ij} \rangle_J^2 =
  \frac{J^2}{N}, \qquad \langle J^{\prime\,2}_{ij} \rangle_J - \langle
  J^{\prime}_{ij} \rangle_J^2 = \frac{J^{\prime\,2}}{N}.
\end{equation}
The mean intrasublattice interactions will always assumed to be
ferromagnetic ($ J^{\prime}_0 > 0$), whereas the mean intersublattice
interactions may be ferromagnetic ($J_0 > 0$) or antiferromagnetic
($J_0 < 0$).

The standard approach to compute the quenched average is to introduce
$n$ non-interacting replicas $\alpha=1,2,\ldots,n$ of the system,
calculate the annealed averages and then take the limit $n \rightarrow
0$ \cite{binder86,fischer91}. In this replica method the free energy
per spin $f$ is given by
\begin{equation}
  f =  \frac{1}{2\beta}\lim_{n \rightarrow 0} \frac{1}{n} \phi, \qquad
  \phi=-\lim_{N \rightarrow \infty} \frac{1}{N} \ln \; 
  \left\langle Z^n \right\rangle_J ,
  \label{replica-trick}
\end{equation}
where $\beta = 1/k_{\mathrm{B}} T$ and $Z^n$ is the partition function
of $n$ replicas of the system.  Performing the average of $Z^n$ over
the random couplings we find
\begin{eqnarray}
  \left\langle Z^n \right\rangle_J&=&\text{Tr} \; \exp -N\left\{
    -\frac{\beta^2 J^2 n}{2} + 
    \beta J^{\prime}_0\frac{n}{N} - \frac{\beta^2
      J^{\prime \, 2}n}{2}\left(1-\frac{n}{N}\right) 
    - \beta H \sum_\alpha \left( m_A^\alpha +
      m_B^\alpha \right) \right. \nonumber \\ &&\mbox{} - \beta J_0
  \sum_\alpha m_A^\alpha m_B^\alpha - \frac{\beta J_0^{\prime}}{2}
  \sum_\alpha \left[ \left(m_A^\alpha \right)^2 +
    \left(m_B^\alpha\right)^2 \right] - \beta^2 J^2 \sum_{(\alpha \beta)}
  q_A^{\alpha \beta} q_B^{\alpha \beta} \nonumber \\ &&\mbox{}
  \left. -\frac{\beta^2 J^{\prime \, 2}}{2} \sum_{(\alpha \beta)} \left[
      \left(q_A^{\alpha \beta} \right)^2 + \left(q_B^{\alpha \beta}
      \right)^2 \right] \right\},
  \label{Zn-averaged}
\end{eqnarray}
where $(\alpha \beta)$ denotes distinct pairs of replicas and we have
introduced the sublattice magnetization and sublattice overlap
function of the replicas,
\begin{equation}
  m_X^\alpha=\frac{1}{N}\sum_{i\in X}S_i^\alpha, \qquad
  q_X^{\alpha\beta}=\frac{1}{N}\sum_{i\in X}S_i^\alpha S_i^\beta,
  \qquad (X=A,B).
  \label{spin-functions}
\end{equation}
The trace over the spin variables in (\ref{Zn-averaged}) can be
performed by taking into account the constraints
(\ref{spin-functions}) by means of the identities
\begin{equation}
  1=\int_{-\infty}^{\infty} d m_X^\alpha
  \int_{-i\infty}^{i\infty} \frac{N d \lambda_X^\alpha}{2 \pi i}
  \exp \left[ -N \lambda_X^\alpha \left( m_X^\alpha-
      \frac{1}{N}\sum_{i\in X}S_i^\alpha \right)\right] \qquad (X=A,B),
\end{equation}
and
\begin{equation}
  1 =\int_{-\infty}^{\infty} d q_X^{\alpha \beta}
  \int_{-i\infty}^{i\infty} \frac{N d \lambda_X^{\alpha\beta}}{2 \pi
    i} \exp \left[ -N \lambda_X^{\alpha\beta} \left(
      q_X^{\alpha\beta}- \frac{1}{N}\sum_{i\in X}S_i^\alpha S_i^\beta
    \right) \right] \qquad (X=A,B).
\end{equation}
We then obtain
\begin{eqnarray}
  \left\langle Z^n \right\rangle_J &=& \prod_\alpha
  \int_{-\infty}^{\infty} d m_A^\alpha \int_{-i\infty}^{i\infty} \frac{N
    d \lambda_A^\alpha}{2 \pi i} \int_{-\infty}^{\infty} d m_B^\alpha
  \int_{-i\infty}^{i\infty} \frac{N d \lambda_B^\alpha}{2 \pi i}
  \prod_{(\alpha \beta)} \int_{-\infty}^{\infty} d q_A^{\alpha \beta}
  \int_{-i\infty}^{i\infty} \frac{N d \lambda_A^{\alpha \beta} }{2 \pi
    i} \nonumber \\ && \mbox{} \times \int_{-\infty}^{\infty} d
  q_B^{\alpha \beta} \int_{-i\infty}^{i\infty} \frac{N d
    \lambda_B^{\alpha \beta} }{2 \pi i} \exp \left[ -N \phi (m_A^\alpha,
    m_B^\alpha,q_A^{\alpha \beta}, q_B^{\alpha \beta}; \lambda_A^\alpha,
    \lambda_B^\alpha,\lambda_A^{\alpha \beta}, \lambda_B^{\alpha \beta})
  \right],
  \label{Zn-traced}
\end{eqnarray}
where
\begin{eqnarray}
  \phi &=& -\frac{\beta^2 J^2 n}{2} + \beta J^{\prime}_0\frac{n}{N} -
  \frac{\beta^2 J^{\prime \, 2}n}{2}\left(1-\frac{n}{N}\right) - \beta H
  \sum_\alpha \left( m_A^\alpha + m_B^\alpha \right) -\beta J_0
  \sum_\alpha m_A^\alpha m_B^\alpha \nonumber \\ &&\mbox{}- \frac{\beta
    J_0^{\prime}}{2} \sum_\alpha \left[ \left(m_A^\alpha \right)^2 +
    \left(m_B^\alpha\right)^2 \right] - \beta^2 J^2 \sum_{(\alpha \beta)}
  q_A^{\alpha \beta} q_B^{\alpha \beta} -\frac{\beta^2 J^{\prime \,
      2}}{2} \sum_{(\alpha \beta)} \left[ \left(q_A^{\alpha \beta} \right)^2
    + \left(q_B^{\alpha \beta} \right)^2 \right] \nonumber \\ &&\mbox{} +
  \sum_{\alpha} \left( \lambda_A^\alpha m_A^\alpha + \lambda_B^\alpha
    m_B^\alpha \right) + \sum_{(\alpha \beta)} \left( \lambda_A^{\alpha
      \beta} q_A^{\alpha \beta} + \lambda_B^{\alpha \beta} q_B^{\alpha
      \beta} \right) - \ln \text{Tr} \; \exp \overline{\mathcal{H}}_A - \ln
  \text{Tr} \; \exp \overline{\mathcal{H}}_B ,
  \label{phi-in-terms-of-lambdas}
\end{eqnarray}
with $\overline{\mathcal{H}}_A$ and $\overline{\mathcal{H}}_B$
denoting the ``effective sublattice Hamiltonians''
\begin{equation}
  \overline{\mathcal{H}}_X = \sum_\alpha \lambda_X^\alpha
  S^\alpha + \sum_{(\alpha \beta)} \lambda_X^{\alpha \beta}
  S^\alpha S^\beta \qquad (X=A,B).
  \label{effective-hamiltonians-in-terms-of-lambdas}
\end{equation}
In the limit of large $N$ the integrations over the $\lambda$
variables in (\ref{Zn-traced}) can be performed by the saddle-point
method. The saddle point is given by 
\begin{equation}
  m_{X}^\alpha = \frac{ \text{Tr}\; S^\alpha \exp
    \overline{\mathcal{H}}_{X} }{ \text{Tr} \;\exp
    \overline{\mathcal{H}}_{X}}, \qquad q_{X}^{\alpha \beta} = \frac{
    \text{Tr}\; S^\alpha S^\beta \exp \overline{\mathcal{H}}_{X} }{
    \text{Tr} \;\exp \overline{\mathcal{H}}_{X}},\qquad (X=A,B).
  \label{saddle-point-equations}
\end{equation}
These equations determine $\lambda$ variables in terms of  $m$
and $q$ variables. The remaining integrations over the $m$ and $q$
variables in (\ref{Zn-traced}) can be performed by the Laplace method
in the limit of large $N$. The stationary-point equations are given by
\begin{equation}
  \lambda_X^\alpha = \beta H +\beta J_0^{\prime} m_X^\alpha + \beta J_0
  m_{\overline{X}}^\alpha , \quad \lambda_X^{\alpha \beta} = \beta^2
  J^{\prime \; 2} q_X^{\alpha \beta} + \beta^2 J^2
  q_{\overline{X}}^{\alpha \beta},\quad (X=A, B),
  \label{stationary-point-equations}
\end{equation}
where $\overline{X}$ is the sublattice complementary to $X$, i.e.,
if $X=A$ then $\overline{X}=B$, and vice versa.  Substituting these
results in the expression of $\phi$ given by Eq.
(\ref{phi-in-terms-of-lambdas}) we find
\begin{eqnarray}
  \phi &=&
  -\frac{\beta^2 n}{2}(J^2 + J^{\prime \, 2})
  +\beta
  J_0 \sum_\alpha m_A^\alpha m_B^\alpha
  +\frac{\beta J_0^{\prime}}{2} \sum_\alpha \left[ \left(m_A^\alpha
    \right)^2 + \left(m_B^\alpha\right)^2 \right]
  + \beta^2 J^2
  \sum_{(\alpha \beta)} q_A^{\alpha \beta} q_B^{\alpha \beta}
  \nonumber \\
  && \mbox{} +\frac{\beta^2 J^{\prime \, 2}}{2} \sum_{(\alpha \beta)}
  \left[ \left(q_A^{\alpha \beta} \right)^2 + \left(q_B^{\alpha \beta}
    \right)^2 \right] - \ln \text{Tr} \; \exp \overline{\mathcal{H}}_A -
  \ln \text{Tr} \; \exp \overline{\mathcal{H}}_B ,
  \label{phi}
\end{eqnarray}
where we have discarded terms that vanish in the limit of large
$N$. Analogously, the effective sublattice Hamiltonians 
(\ref{effective-hamiltonians-in-terms-of-lambdas}) become
\begin{equation}
  \overline{\mathcal{H}}_X = \beta \sum_\alpha \left( H + J_0^{\prime}
    m_X^\alpha +J_0 m_{\overline{X}}^\alpha \right) S^\alpha + \beta^2
  \sum_{(\alpha \beta)} \left( J^{\prime \; 2} q_X^{\alpha \beta} + J^2
    q_{\overline{X}}^{\alpha \beta} \right) S^\alpha S^\beta, \quad
  (X=A,B).
\end{equation}

To evaluate the general expressions obtained thus far it is necessary
to impose some structure on $m$ and $q$ variables.  The simplest
assumption corresponds to the RS solution\cite{binder86,fischer91}
obtained by assuming order parameters independent of replica indices,
\begin{equation}
  m_X^\alpha = m_X , \qquad q_X^{\alpha\beta} = q_X, \qquad (X=A,B).
\end{equation}
Proceeding in the usual way \cite{binder86,fischer91}, one finds that
the saddle-point equations (\ref{saddle-point-equations}) and
stationary-point equations (\ref{stationary-point-equations}) give the
equations of state
\begin{equation}
  m_X = \left\langle \tanh H_X \right\rangle , \qquad q_X =
  \left\langle \tanh^2 H_X \right\rangle, \qquad (X=A,B),
  \label{RS-order-parameters}
\end{equation}
where
\begin{equation}
  H_X = \beta \left( H + J_0^\prime m_X + J_0 m_{\overline{X}} + \sqrt{
      J^{\prime \; 2} q_X + J^2 q_{\overline{X}} } \; x \right), \qquad
  (X=A,B),
\end{equation}
and the brackets without subscript $\langle \cdots \rangle$ denote
Gaussian averages,
\begin{equation}
  \langle \cdots \rangle = \int_{-\infty}^{\infty} \frac{d x}{\sqrt{2
      \pi} } e^{-x^2/2} ( \cdots ).
\end{equation}
Analogously, the free energy per spin (\ref{replica-trick}) becomes
\begin{eqnarray}
  f &=& -\frac{\beta J^2}{4} \left( 1-q_A \right) \left( 1-q_B \right)
  -\frac{\beta J^{\prime \; 2}}{8} \left[ \left( 1-q_A \right)^2 +
    \left( 1-q_B \right)^2 \right] + \frac{J_0}{2} m_A m_B +
  \frac{J_0^\prime}{4} \left( m_A^2 + m_B^2 \right) \nonumber \\ &&
  \mbox{} - \frac{1}{2 \beta} \left\langle \ln 2 \cosh H_A \right\rangle
  - \frac{1}{2 \beta} \left\langle \ln 2 \cosh H_B\right\rangle.
  \label{RS-free-energy}
\end{eqnarray}


\section{The stability of replica-symmetric solution}

The validity of the RS solution (\ref{RS-order-parameters}) rests on
the applicability of Laplace method used to perform the integrations
over $m$ and $q$ variables for large $N$. The integral converges only
if the stationary point (\ref{stationary-point-equations}) is a
minimum of $\phi$, i.e., only if the eigenvalues of the Hessian matrix
formed by the second derivatives of the function $\phi$ given by
equation (\ref{phi-in-terms-of-lambdas}) with respect to the $m$ and
$q$ variables are all positive.  We can equivalently consider $\phi$
as a function of $\lambda$ variables, related to $m$ and $q$ variables
by means of Eq.  (\ref{saddle-point-equations}). We will follow the
latter approach because it leads to simpler calculations.  The Hessian
is a $[n(n+1)/2] \times [n(n+1)/2]$ matrix whose elements are $2
\times 2$ matrices given by
\begin{equation}
  \mathbf{G}^{\alpha \beta}=\left(
    \begin{array}{cc}
      G^{\alpha \beta}_{A A}&
      G^{\alpha \beta}_{A B} \\
      G^{\alpha \beta}_{B A}&
      G^{\alpha \beta}_{B B}
    \end{array}
  \right), 
  \mathbf{G}^{\alpha (\beta\gamma)}=\left(
    \begin{array}{cc}
      G^{\alpha (\beta\gamma)}_{A A} &
      G^{\alpha (\beta\gamma)}_{A B} \\
      G^{\alpha (\beta\gamma)}_{B A} &
      G^{\alpha (\beta\gamma)}_{B B} 
    \end{array}
  \right), 
  \mathbf{G}^{(\alpha\beta) {(\gamma\delta)}}=\left(
    \begin{array}{cc}
      G^{(\alpha\beta) (\gamma\delta)}_{A A} &
      G^{(\alpha\beta) (\gamma\delta)}_{A B} \\
      G^{(\alpha\beta) (\gamma\delta)}_{B A} &
      G^{(\alpha\beta) (\gamma\delta)}_{B B}
    \end{array}
  \right),
  \label{hessian-matrix}
\end{equation}
where
\begin{equation}
  G^{\alpha \beta}_{X Y}=\frac{\partial^2 \phi}{\partial
    \lambda_X^{\alpha} \partial \lambda_Y^{\beta}}, \quad G^{\alpha
    (\beta\gamma)}_{X Y}=\frac{\partial^2 \phi}{\partial
    \lambda_X^{\alpha} \partial \lambda_Y^{(\beta\gamma)}}, \quad
  G^{(\alpha\beta) (\gamma\delta)}_{X Y}=\frac{\partial^2 \phi}{\partial
    \lambda_X^{\alpha\beta} \partial \lambda_Y^{\gamma\delta}} \qquad
  (X,Y=A,B).
\end{equation}
At the stationary point of the RS solution
(\ref{RS-order-parameters}) there are seven different types of $2
\times 2$ elements of the Hessian matrix. We denote these elements by
\cite{almeida78}
\begin{equation}
  \begin{array}{lllll}
    \mathbf{G}^{\alpha\alpha}=\mathbf{A}, \qquad &
    \mathbf{G}^{\alpha\beta}=\mathbf{B}, \qquad &
    \mathbf{G}^{\alpha(\alpha\beta)}=\mathbf{C}, \qquad &
    \mathbf{G}^{(\alpha\beta)\alpha}=\mathbf{\widetilde{C}}, \qquad
    \mathbf{G}^{\alpha(\beta\gamma)}=\mathbf{D}, \qquad \\
    \mathbf{G}^{(\alpha\beta)\gamma}=\mathbf{\widetilde{D}}, \qquad &
    \mathbf{G}^{(\alpha\beta)(\alpha\beta)}=\mathbf{P}, \qquad &
    \mathbf{G}^{(\alpha\beta)(\alpha\gamma)}=\mathbf{Q}, \qquad &
    \mathbf{G}^{(\alpha\beta)(\gamma\delta)}=\mathbf{R},
  \end{array}
  \label{hessian-matrix-elements}
\end{equation}
where the indices $\alpha$, $\beta$, $\gamma$ and $\delta$ are all distinct
and the tilde denotes the transpose of the matrix. We do not quote the 
lengthy expressions for these elements because only their linear
combinations are needed in the calculation of the eigenvalues.

The eigenvalues of the Hessian matrix can now be determined by finding
the eigenvectors that divide the space into orthogonal subspaces
closed to the permutation operation. The procedure are analogous to
the case of the SK model \cite{almeida78} except that now the elements
of the Hessian matrix are $2 \times 2$ matrices
(\ref{hessian-matrix-elements}). These eigenvectors are
\cite{dominicis85}: $n(n-3)$ transversal or replicon eigenvectors
depending on two replica indices, $4 (n-1)$ anomalous eigenvalues
depending on a single replica index, and 4 longitudinal eigenvectors
independent of replica indices.

The eigenvalues associated with the transversal eigenvectors are found
to be the eigenvalues of the $2 \times 2$ matrix
\begin{equation}
  \mathbf{T}=\mathbf{P}-2\mathbf{Q}+\mathbf{R},
\end{equation}
with elements 
\begin{eqnarray}
  T_{11}&=&(1-2q_A+r_A)-(\beta J^{\prime})^2(1-2q_A+r_A)^2, \\
  T_{12}&=&T_{21}=-(\beta J)^2 (1-2q_A+r_A)(1-2q_B+r_B), \\
  T_{22}&=&(1-2q_B+r_B)-(\beta J^{\prime})^2(1-2q_B+r_B)^2,
\end{eqnarray}
where
\begin{equation}
  t_X = \left\langle \tanh^3 H_X \right\rangle , \qquad r_X =
  \left\langle \tanh^4 H_X \right\rangle, \qquad (X=A,B).
\end{equation}
The necessary and sufficient condition for all the eigenvalues to be
positive are
\begin{equation}
  T_{11} + T_{22} > 0 \qquad \text{and} \qquad
  T_{11}T_{22}-T_{12}^2 > 0,
\end{equation}
which are equivalent to the conditions
\begin{eqnarray}
  T_1&=&2-(\beta J^{\prime})^2(1-2q_A+r_A)-(\beta J^{\prime})^2(1-2q_B+r_B)
  > 0,
  \label{T1}\\
  T_2&=&[1-(\beta J^{\prime})^2(1-2q_A+r_A)][1-(\beta J^{\prime})^2(1-2q_B+r_B)]
  \nonumber \\
  && \mbox{} - (\beta J)^4(1-2q_A+r_A)(1-2q_B+r_B) > 0,
  \label{T2}
\end{eqnarray}
in agreement with previous studies \cite{korenblit85,takayama88}. A RS
solution satisfying these conditions will be called transversally (T)
stable, and T unstable otherwise.

The eigenvalues associated with anomalous and longitudinal
eigenvectors are the same in the limit $n \rightarrow 0$. They are
found to be the eigenvalues of the $4 \times 4$ matrix
\begin{equation}
  \mathbf{L}=\left(
    \begin{array}{ll}
      \mathbf{A}-\mathbf{B} & \mathbf{D}-\mathbf{C} \\
      2\mathbf{\widetilde{C}}-2\mathbf{\widetilde{D}} &
      \mathbf{P}-4\mathbf{Q}+3\mathbf{R}
    \end{array}
  \right),
\end{equation}
where
\begin{eqnarray}
  L_{11}&=&(1-q_A)-\beta J_0^{\prime}(1-q_A)^2+2(\beta
  J^{\prime})^2(t_A-m_A)^2,\\ 
  L_{22}&=&(1-q_B)-\beta J_0^{\prime}(1-q_B)^2+2(\beta
  J^{\prime})^2(t_B-m_B)^2,\\ 
  L_{12}&=&L_{21}=-\beta
  J_0(1-q_A)(1-q_B)+2 (\beta J)^2(t_A-m_A)(t_B-m_B), \\
  L_{13}&=&-\frac{1}{2}L_{31}=(t_A-m_A)[1-\beta J_0^{\prime}(1-q_A)
  -(\beta J^{\prime})^2(1-4q_A+3r_A)], \\
  L_{24}&=&-\frac{1}{2}L_{42}=(t_B-m_B)[1-\beta J_0^{\prime}(1-q_B)
  -(\beta J^{\prime})^2(1-4q_B+3r_B)], \\
  L_{14}&=&-\frac{1}{2}L_{41}=-\beta J_0(t_B-m_B)(1-q_A)-
  (\beta J)^2 (t_A-m_A)(1-4q_B+3r_B), \\
  L_{23}&=&-\frac{1}{2}L_{32}=-\beta J_0(t_A-m_A)(1-q_B)-
  (\beta J)^2 (t_B-m_B)(1-4q_A+3r_A), \\
  L_{33}&=&(1-4q_A+3r_A)[1-(\beta J^{\prime})^2(1-4q_A+3r_A)]+2 \beta
  J_0^{\prime}(t_A-m_A)^2, \\
  L_{44}&=&(1-4q_B+3r_B)[1-(\beta J^{\prime})^2(1-4q_B+3r_B)]+2 \beta
  J_0^{\prime}(t_B-m_B)^2, \\
  L_{34}&=&L_{43}=-(\beta J)^2 (1-4q_A+3r_A)(1-4q_B+3r_B)
  +2 \beta J_0(t_A-m_A)(t_B-m_B).
\end{eqnarray}
The characteristic equation has the form
\begin{equation}
  \lambda^4-a_1 \lambda^3+a_2 \lambda^2-a_3 \lambda+a_4=0,
  \label{characteristc-equation}
\end{equation}
where the coefficients $a_n$ are $n$-th order traces of the matrix
$\mathbf{L}$.  A numerical study of equation
(\ref{characteristc-equation}) shows that the eigenvalues are complex
for some values of the parameters of the model, in contrast with
one-sublattice SK model in which the anomalous and longitudinal
eigenvalues never become complex \cite{almeida78}.  Even
though the Hessian matrix (\ref{hessian-matrix}) for $n > 1$ is real
and symmetric, in the limit $n \rightarrow 0$ there is no guarantee
that the eigenvalues will be real.  In fact, complex longitudinal and
anomalous eigenvalues also arise in the spin 1 one-sublattice
infinite-range spin-glass model with crystal-field anisotropy
\cite{lage82,costa94a}.  In general, therefore, the stability
condition should require the real part of the eigenvalues to be
positive. According to the Hurwitz criterion \cite{uspensky48}, the
necessary and sufficient condition for all the roots of equation
(\ref{characteristc-equation}) to have positive real parts are
\begin{eqnarray}
  D_1&=&a_1 > 0 ,\qquad
  D_2=\left|
    \begin{array}{cc}
      a_1 & a_3 \\
      1  & a_2
    \end{array}
  \right| = a_1a_2-a_3> 0, \\
  D_3&=&\left|
    \begin{array}{ccc}
      a_1 & a_3 & 0 \\
      1   & a_2 & a_4 \\
      0   & a_1 & a_3
    \end{array}
  \right| = a_3 D_2 - a_1^2a_4 > 0, \qquad
  D_4=\left|
    \begin{array}{cccc}
      a_1 & a_3 & 0   & 0\\
      1   & a_2 & a_4 & 0\\
      0   & a_1 & a_3 & 0\\
      0   & 1   & a_2 &a_4 
    \end{array}
  \right| = a_4 D_3 > 0. 
\end{eqnarray}
These condition are equivalent to the following four conditions:
\begin{eqnarray}
  L_1 &=& a_1  >  0, \label{L1} \\
  L_2 &=& D_2  =  a_1 a_2 -a_3 > 0,  \label{L2} \\
  L_3 &=& D_3  =  a_1a_2a_3 - a_3^2  - a_4 a_1^2 > 0,  \label{L3}\\
  L_4 &=& a_4  >  0. \label{L4}
\end{eqnarray}
A RS solution satisfying these conditions will be called
longitudinally (L) stable, and L unstable otherwise.


\section{Results of the stability analysis}

In this section we present the results of the stability analysis of
the RS solution for different values of the parameters of the model.
Since the Hamiltonian (\ref{hamiltonian}) is invariant under the
simultaneous transformations
\begin{equation}
  H \rightarrow -H, \qquad S_i \rightarrow -S_i,
\end{equation}
it is sufficient to consider fields $H \geq 0$. For $H=0$ only one of
the two solutions related by the global inversion symmetry has to be
considered. 

\subsection{Zero applied  field}

\subsubsection{Ferromagnetic intersublattice interaction}

\begin{figure}[tbp]
  \begin{center}
    \includegraphics[height=14cm]{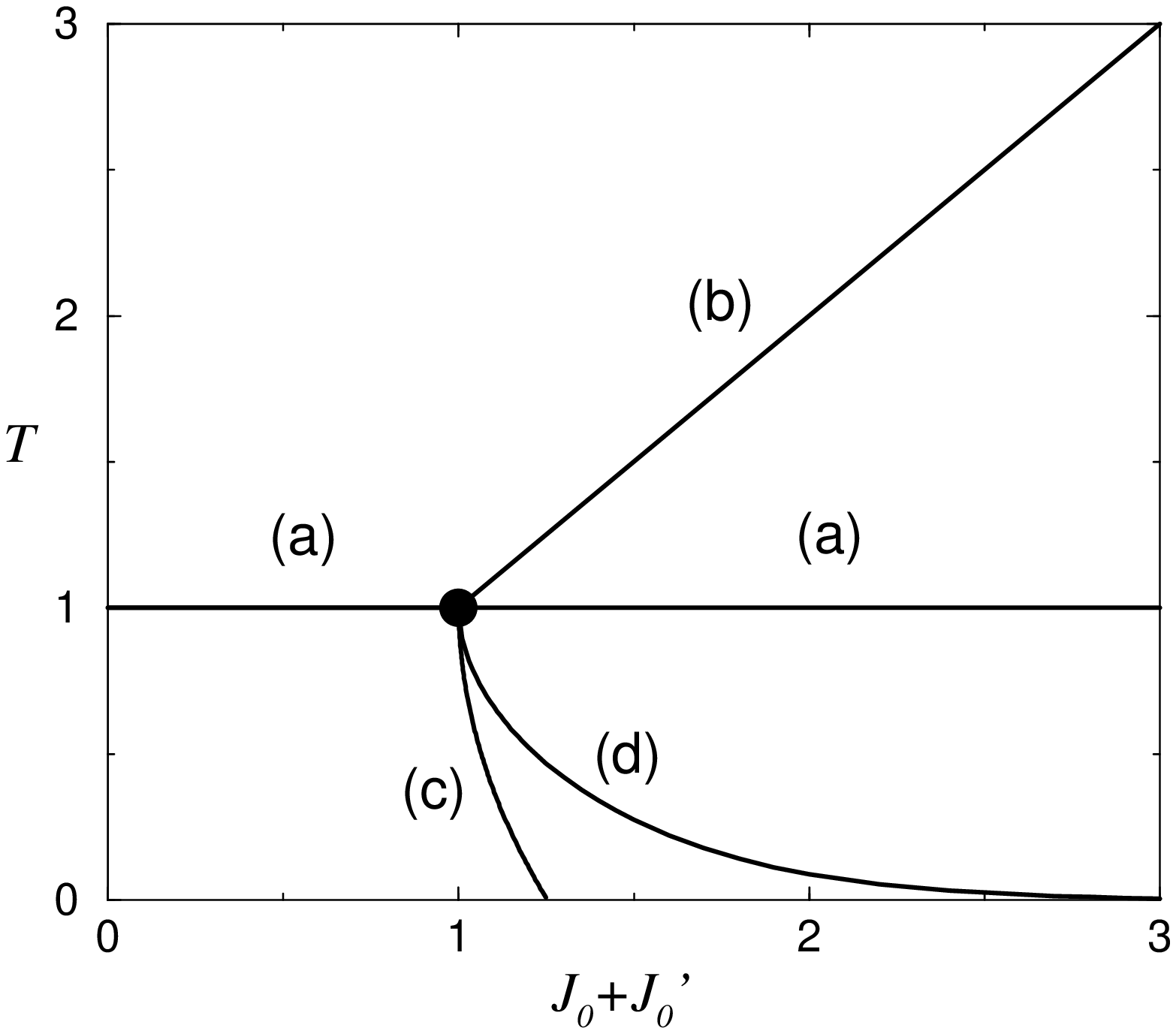}
  \end{center}
  \protect \caption{Regions of the zero-field phase diagram where
    different types of solution are possible. For ferromagnetic
    intersublattice interaction ($J_0>0$), the P solution is L stable
    above line (b) and the left portion of line (a), the F solution
    between lines (b) and (c), and SG solution between the left portion of
    line (a) and line (c). The P solution is T stable above line (a), the
    F solution between lines (b) and (d). The SG solution is T unstable
    between the left side of line (a) and line (c), and F solution between
    lines (c) and (d). For antiferromagnetic intersublattice interaction
    ($J_0<0$) the F solution is replaced by AF solution and the label of
    horizontal axis by $-J_0+J_0^\prime$. The temperature and energy units
    in the axis are such that $\protect\sqrt{J^2+J^{\prime\,2}}=1$ and
    $k_B=1$}
  \label{fig1}
\end{figure}

In zero applied field ($H = 0$) and ferromagnetic intersublattice
interactions ($J_0 > 0$) the solutions of the set of equations
(\ref{RS-order-parameters}) are of the form
\begin{equation}
  m_A=m_B=m, \qquad q_A=q_B=q.
\end{equation}
Three types of solutions are possible:
\begin{itemize}

\item Paramagnetic (P) solution: $m=0, q=0$. 

\item Spin Glass (SG) solution: $q>0, m=0$.

\item Ferromagnetic (F) solution: $q>0, m>0$.

\end{itemize}
Fig. \ref{fig1} shows the lines delimiting the regions where
different types of solutions can be found in the plane of temperature
versus $J_0 + J_0^\prime$. 

The P solution is always possible. However it is L stable only above
the line (b) and the left portion of line (a), and T stable above line
(a). The L instability of P solution occurs due to the violation of
the condition (\ref{L4}), which is given in the case of P solution by
\begin{equation}
  L_4 = [1-\beta^2(J^{\prime\,2}-J^2)] [1-\beta^2(J^{\prime\,2}+J^2)]
  [1-\beta(J_0^\prime-J_0)] [1-\beta(J_0^\prime+J_0)] > 0.
  \label{L4P}
\end{equation}
For $(J_0 + J_0^\prime)/\sqrt{J^2+J^{\prime\,2}} \leq 1/2$, the second
factor in (\ref{L4P}) becomes negative below line (a). Thus the
left portion of line (a) is determined by
\begin{equation}
  \beta^2(J^{\prime\,2}+J^2)=1.
  \label{line-a}
\end{equation}
On the other hand, for $(J_0 + J_0^\prime)/\sqrt{J^2+J^{\prime\,2}}>
1/2$ the fourth factor in (\ref{L4P}) becomes negative below line
(b).  Thus the equation for line (b) is
\begin{equation}
  \beta(J_0^\prime+J_0)=1.
  \label{line-b}
\end{equation}
The T instability of the P solution is due to the violation of the
condition (\ref{T2}), which is given in the case of P solution  by
\begin{equation}
  T_2 = [1-\beta^2(J^{\prime\,2}-J^2)] [1-\beta^2(J^{\prime\,2}+J^2)] > 0.
  \label{T2P}
\end{equation}
The second factor in (\ref{T2P}) becomes negative below line (a) for
all values of $J_0+J_0^\prime$. Thus line (a) is given by equation
(\ref{line-a}) for all $J_0+J_0^\prime$. The T and L instabilities of
the P solution occur simultaneously on the line (a) for $ (J_0 +
J_0^\prime)/\sqrt{J^2+J^{\prime\,2}} \leq 1/2$.

The SG solution is possible only below line (a). It is T unstable
throughout this region and L stable to the left of line (c).  The L
instability of the SG solution occurs due to the violation of the
condition (\ref{L4}), which is given in the case of SG solution by
\begin{eqnarray}
  L_4= (1-q)^2(1-4q+3r)^2 && [1-\beta^2(J^{\prime\,2}-J^2)(1-4q+3r)]
  [1-\beta^2(J^{\prime\,2}+J^2)(1-4q+3r)] \nonumber \\ && \mbox{} \times
  [1-\beta(J_0^\prime-J_0)(1-q)] [1-\beta(J_0^\prime+J_0)(1-q)] > 0.
  \label{L4SG}
\end{eqnarray}
For $(J_0 + J_0^\prime)/\sqrt{J^2+J^{\prime\,2}} > 1/2$ the last
factor in (\ref{L4SG}) becomes negative to the left of line (c). Thus
the equation determining line (c) is
\begin{equation}
  \beta(J_0^\prime+J_0)(1-q)=1.
\end{equation}

The F solution is possible only between lines (b) and (c). It is L
stable throughout this region but T stable only above line (d).  The T
instability of the F solution occurs due to the violation of the
condition (\ref{T2}), which is given in the case of F solution by
\begin{equation}
  T_2=[1-\beta^2(J^{\prime\,2}-J^2)(1-2q+3r)]
  [1-\beta^2(J^{\prime\,2}+J^2)(1-2q+3r)] > 0.
  \label{T2F}
\end{equation}
For $(J_0 + J_0^\prime)/\sqrt{J^2+J^{\prime\,2}} > 1/2$ the second
factor in (\ref{T2F}) becomes negative below line (d). Thus line (d)
is described by equation
\begin{equation}
  \beta^2(J^{\prime\,2}+J^2)(1-2q+3r)=1.
  \label{line-d}
\end{equation}

Rejecting solutions that are L unstable, we conclude that the P phase
is located above lines (b) and left portion of line (a), the SG phase
between the left portion of line (a) and line (c), and finally the F
phase between lines (b) and (c).  The SG phase, and the F phase
between lines (c) and (d), are T unstable, indicating the need for a
replica-symmetry-breaking solution in this region. The transition line
(c) will change to a vertical line if such a solution is considered
\cite{toulouse80}.  We mention that, as should be expected, in the
case $J_0 =0$ and $J=0$, or $J_0^\prime=0$ and $J^\prime=0$, these
results reduce to those of one-sublattice SK model
\cite{binder86,fischer91}.

\subsubsection{Antiferromagnetic intersublattice interaction}

The Hamiltonian (\ref{hamiltonian}) in zero applied field ($H = 0$) is
invariant under simultaneous transformations
\begin{equation}
  J_0 \longrightarrow -J_0, \qquad S_i \longrightarrow -S_i \qquad (i \in B).
\end{equation}
In fact, we can check explicitly that all the expressions for the RS
solution, including those of stability conditions, are invariant under
simultaneous transformations
\begin{equation}
  J_0 \longrightarrow -J_0, \qquad m_B \longrightarrow -m_B.
\end{equation}
Thus the case of antiferromagnetic intersublattice interaction $J_0<0$
is completely equivalent to the case of ferromagnetic intersublattice
interaction $-J_0>0$ by replacing $m_B$ by $-m_B$. This means that the F
solution is replaced by the antiferromagnetic (AF) solution
\begin{equation}
  m_A=-m_B=m, \qquad q_A=q_B=q.
\end{equation}
The results displayed in Fig. \ref{fig1} remains valid, with $J_0$
replaced by $-J_0$ and F solution by AF solution.


\subsection{Non-zero applied field}

\subsubsection{Ferromagnetic intersublattice interaction}

\begin{figure}[tbp]
  \begin{center}
    \includegraphics[height=14cm]{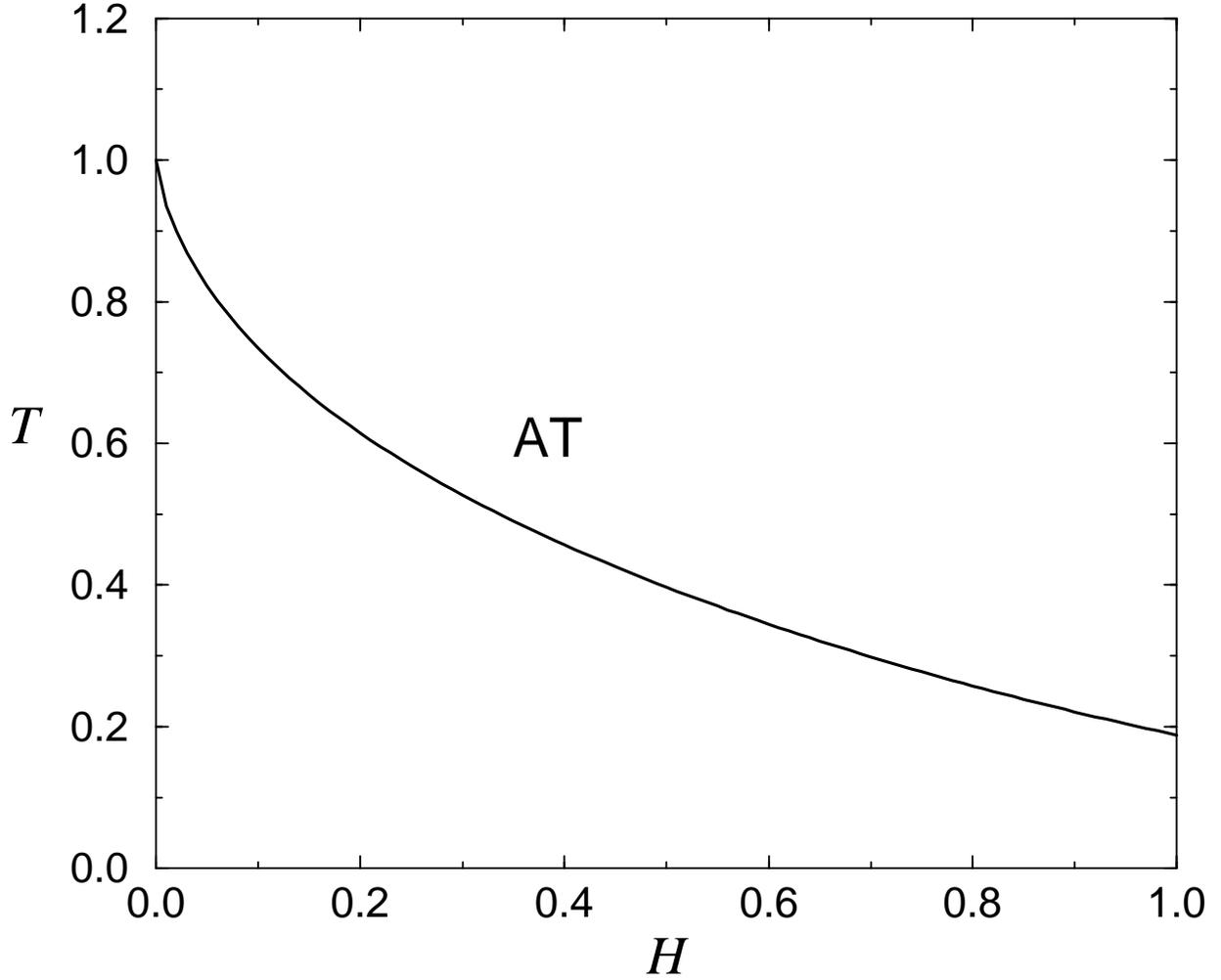}
  \end{center}
  \protect \caption{Region of stability in the presence of a field for
    the case of ferromagnetic intersublattice interaction ($J_0 <0$). The
    values of parameters are $J^\prime/J=1$, $J_0^\prime/J_0=1$ and $(J_0
    + J_0^\prime)/\protect\sqrt{J^2+J^{\prime\,2}}=1/2$. There is only the
    P solution which is always L stable but becomes T unstable below the
    de Almeida-Thouless line AT. The temperature and field units in the
    axis are such that $\protect\sqrt{J^2+J^{\prime\,2}}=1$ and $k_B=1$}
  \label{fig2}
\end{figure}

In non-zero applied field ($H > 0$) and ferromagnetic intersublattice
interactions ($J_0 > 0$), only the paramagnetic (P) solution is
possible for the set of equations (\ref{RS-order-parameters}), which
are of the form
\begin{equation}
  m_A=m_B=m>0, \qquad q_A=q_B=q>0.
\end{equation}
This solution is always L stable, but becomes T unstable for low
temperatures due to the violation of the condition (\ref{T2}), which
in this case is also given by Eq. (\ref{T2F}). The instability line
is given by Eq. (\ref{line-d}), illustrated in Fig. \ref{fig2} for the
case $J^\prime/J=1$, $J_0^\prime/J_0=1/2$ and $(J_0 +
J_0^\prime)/\protect\sqrt{J^2+J^{\prime\,2}}=2$ .  As should be
expected, in the case $J_0 =0$ and $J=0$, or $J_0^\prime=0$ and
$J^\prime=0$, these results reduce to the de Almeida-Thouless line
of one-sublattice SK model\cite{almeida78}.


\subsubsection{Antiferromagnetic intersublattice interaction}

In non-zero applied field ($H > 0$) and antiferromagnetic
intersublattice interactions ($J_0 < 0$), two types of solutions to
the set of equations (\ref{RS-order-parameters}) are possible:
\begin{itemize}

\item Paramagnetic (P) solution: $m_A=m_B=m>0, q_A=q_B=q>0$.

\item Antiferromagnetic (AF) solution: $m_A \ne m_B, q_A \ne q_B$.

\end{itemize}

For $(-J_0 + J_0^\prime)/\sqrt{J^2+J^{\prime\,2}} \leq 1/2$ only P
solution is possible. This solution is always L stable but becomes T
unstable at low temperatures due to the to the violation of the
condition (\ref{T2}), which in this case it is also given by
(\ref{T2F}). The instability line is given by Eq. (\ref{line-d}).

\begin{figure}[tbp]
  \begin{center}
    \includegraphics[height=14cm]{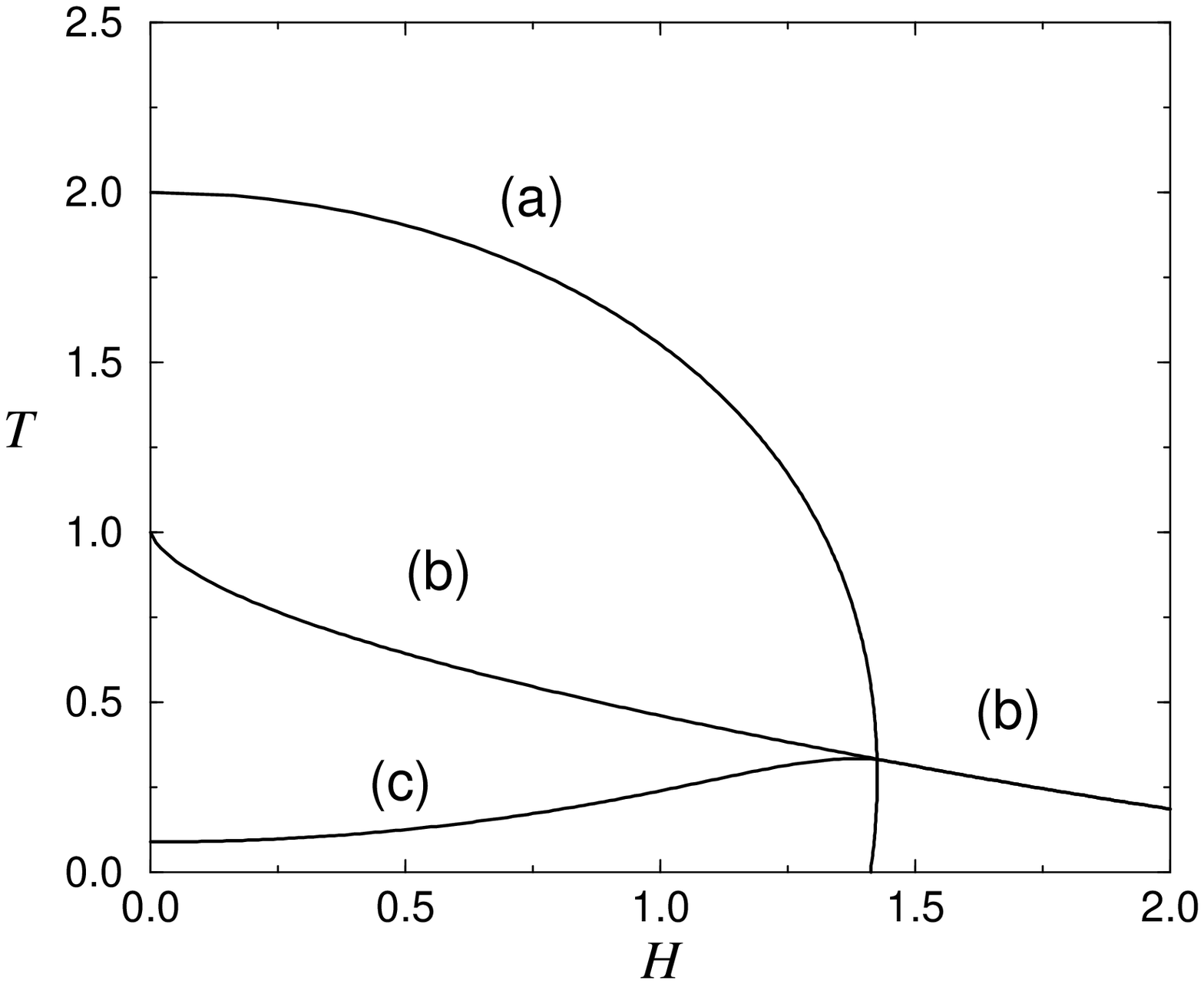}
  \end{center}
  \protect \caption{Regions of stability and existence of different
    solutions in the temperature versus field phase diagram for the case
    of antiferromagnetic intersublattice interaction ($J_0 <0$). The
    values of parameters are $J^\prime/J=1$, $-J_0^\prime/J_0=1/2$ and
    $(-J_0 + J_0^\prime)/\protect\sqrt{J^2+J^{\prime\,2}}=2$. The P
    solution is L stable outside line (a) and T stable above line (b). The
    AF solution is possible only inside line (a) and it is always L
    stable, but becomes T unstable below line (c).  The temperature and
    field units in the axis are such that
    $\protect\sqrt{J^2+J^{\prime\,2}}=1$ and $k_B=1$}
  \label{fig3}
\end{figure}

For $(-J_0 + J_0^\prime)/\sqrt{J^2+J^{\prime\,2}} > 1/2$, AF solution
also becomes possible.  Fig. \ref{fig3} shows the lines delimiting the
regions of existence and stability of each type of solution for
$-J_0^\prime/J_0=1/2$. The P solution becomes L unstable below line (a)
due to the violation of the condition (\ref{L4}). In the case of P
solution this condition is given by
\begin{eqnarray}
  L_4&=&[(1-q)(1-4q+3r)+2(t-m)^2]^2 \left\{
    [1-\beta(J_0^\prime+J_0)(1-q)] [1-\beta (J^{\prime\,2}-J^2 )
  \right. \nonumber \\
  && \mbox{} \left.  \times (1-4q+3r)]+2\beta^3 (J_0^\prime+J_0)
    (J^{\prime\,2}-J^2) (t-m)^2 \right\} \left\{
    [1-\beta(J_0^\prime-J_0)(1-q)] \right. \nonumber \\
  && \mbox{} \times \left.  [1-\beta (J^{\prime\,2}+J^2 )(1-4q+3r)] +
    2\beta^3 (J_0^\prime-J_0) (J^{\prime\,2}+J^2) (t-m)^2 \right\} > 0.
  \label{L4AF}
\end{eqnarray}
The first factor in (\ref{L4AF}) becomes negative inside line (a).
Therefore the equation determining line (a) is
\begin{equation}
  [1-\beta(J_0^\prime-J_0)(1-q)] [1-\beta (J^{\prime\,2}-J^2 )
  (1-4q+3r)]+2\beta^3 (J_0^\prime-J_0) (J^{\prime\,2}-J^2) (t-m)^2 = 0,
  \label{P-AF}
\end{equation}
which is in agreement with previous study \cite{fyodorov87b}.  The P
solution is T unstable below line (b). This instability occurs due to
the violation of condition (\ref{T2}), given in this case by Eq.
(\ref{T2F}), caused by the second factor. Therefore the line (b) is
determined by equation (\ref{line-d}).  The AF solution is possible
only inside line (a). It is L stable throughout this region and T
unstable below line (c). This instability is due to the violation of
condition (\ref{T2}). Therefore line (c) obeys the equation
\begin{equation}
  [1-(\beta J^{\prime})^2(1-2q_A+r_A)][1-(\beta
  J^{\prime})^2(1-2q_B+r_B)] = (\beta J)^4(1-2q_A+r_A)(1-2q_B+r_B).
  \label{line-c-AF}
\end{equation}
Rejecting solutions that are L unstable, we conclude that P phase
exists outside and AF phase inside line (a). The P solution becomes T
unstable below line (b) and AF solution below line (c), which meet
smoothly on the line (a). In the region below lines (b) and (c) it is
necessary to consider replica symmetry breaking solution, which will
presumably change line (a) in this region.

\begin{figure}[tbp]
  \begin{center}
    \includegraphics[height=14cm]{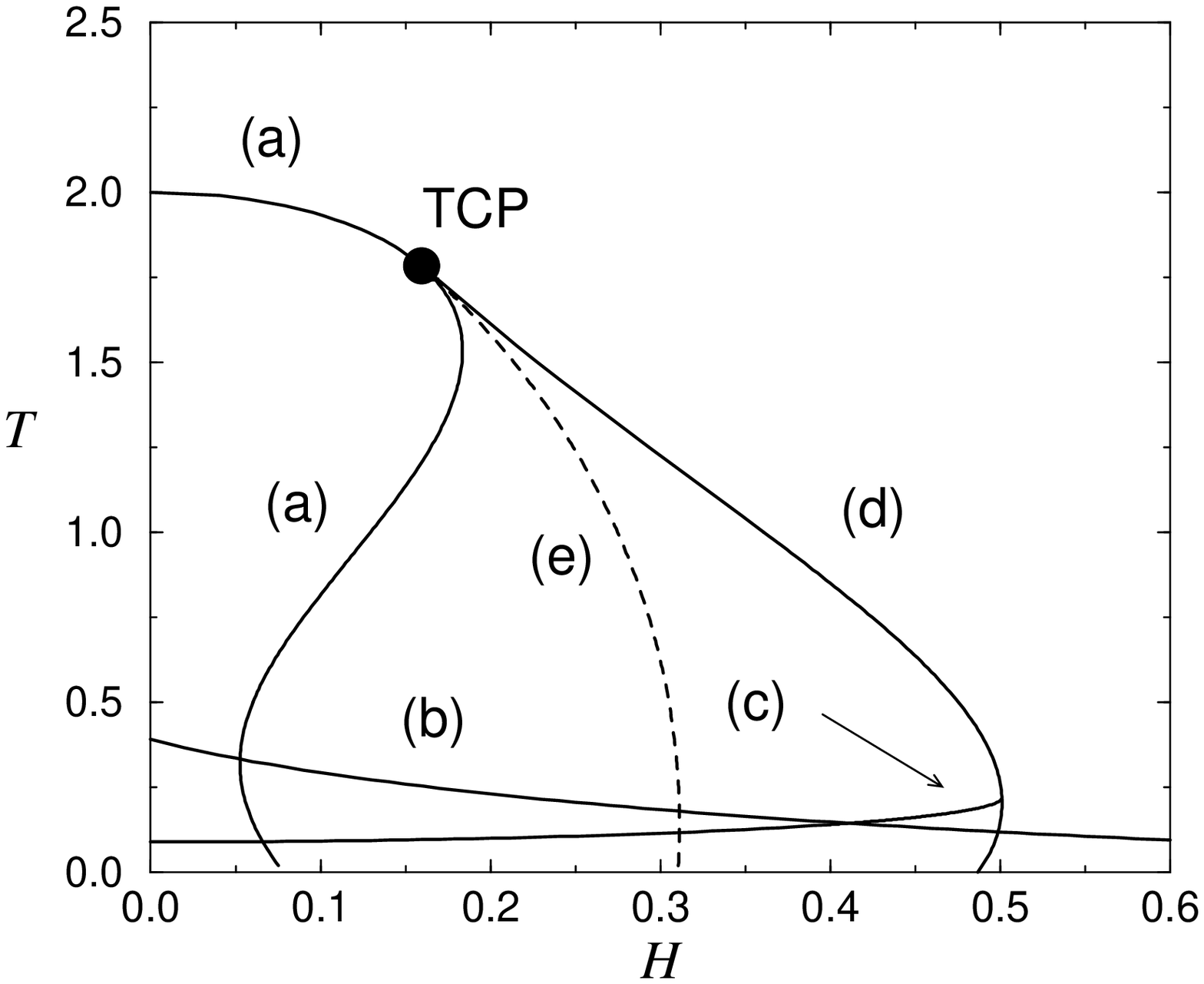}
  \end{center}
  \protect \caption{Regions of stability and existence of different
    solutions in the temperature versus field phase diagram for the case
    of antiferromagnetic intersublattice interaction ($J_0 <0$). The
    values of parameters are $J^\prime/J=1$, $-J_0^\prime/J_0=5$ and
    $(-J_0 + J_0^\prime)/\protect\sqrt{J^2+J^{\prime\,2}}=2$.  The P
    solution is L stable outside the line (a) and T stable above line
    (b). There is one AF solution inside line (a) which is always L
    stable, and two distinct AF solutions between lines (a) and (d), one L
    stable and the other L unstable. The L stable AF solution is also T
    stable above line (c).  The line (e) is the first-order transition
    line and TCP is the tricritical point.  The temperature and field
    units in the axis are such that $\protect\sqrt{J^2+J^{\prime\,2}}=1$
    and $k_B=1$.}
  \label{fig4}
\end{figure}

\begin{figure}[tbp]
  \begin{center}
    \includegraphics[height=14cm]{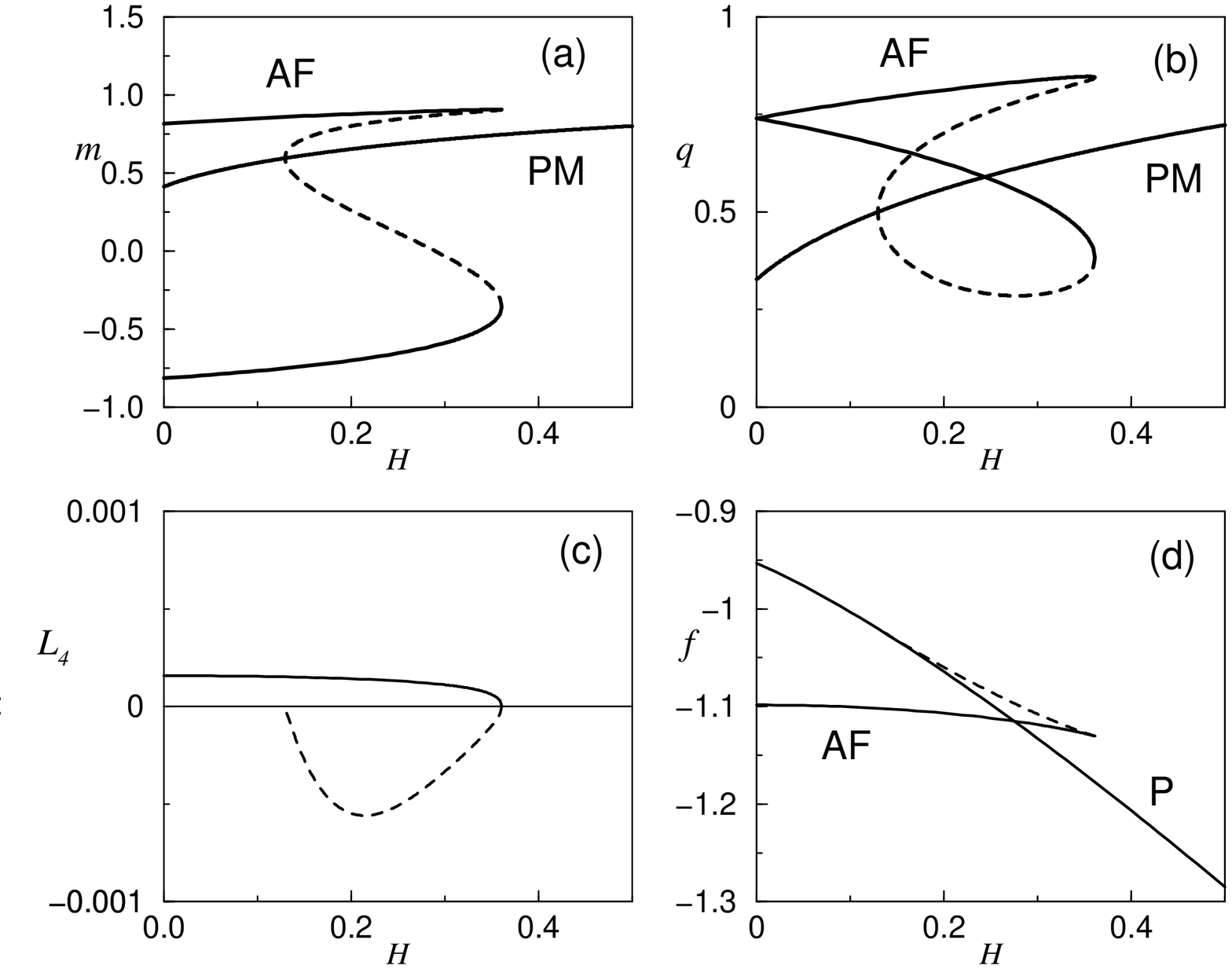}
  \end{center}
  \protect \caption{Field behavior of various quantities for the fixed
    value of temperature $k_BT/\protect\sqrt{J^2+J^{\prime\,2}}=1$ in the
    phase diagram of Fig. \ref{fig4}. In (a) and (b) the order parameters
    of the P and AF solution are shown, in (c) the L stability condition
    $L_4$ of the AF solution, and finally in (d) the free-energy per spin
    of P and AF solutions. The L unstable AF solution are represented by
    dotted lines, and corresponds to the upper portion of the van der
    Waals loop in (d). The first-order transition is determined by the
    intersection of L stable AF and P solutions, as depicted in (d).  The
    temperature and energy units in the axis are such that
    $\protect\sqrt{J^2+J^{\prime\,2}}=1$ and $k_B=1$.}
  \label{fig5}
\end{figure}

For sufficiently large values of the ratio $-J_0^\prime/J_0$ the model
can exhibit first-order transition from AF phase to the P phase
\cite{fyodorov87b}. As an example, we consider the case
$-J_0^\prime/J_0=5$ shown in Fig. \ref{fig4}.  The P solution is L
stable inside line (a) given by Eq. (\ref{P-AF}), and T stable above
line (b) given by Eq. (\ref{line-c-AF}). There is one AF solution
inside line (a) and two distinct AF solutions between lines (a) and
(d), as illustrated in Figs. \ref{fig5}(a) and \ref{fig5}(b) for $k_B
T/\sqrt{J^2+J^{\prime\,2}}=1$. One of the AF solutions, corresponding
to dotted lines in Fig. \ref{fig5}, is L unstable due to the violation
of the condition (\ref{L4}), as shown in Fig. \ref{fig5}(c). The
transition between AF phase and P phase is first order, determined by
equating the free energies of L stable AF and P phases, as shown in
Fig. \ref{fig5}(d). The first-order transition line is shown as dotted
line in Fig. \ref{fig4}, which ends at the tricritical point TCP.  The
L stable AF solution becomes T unstable below line (c) due to the
violation of condition (\ref{T2}), and it is determined by Eq.
(\ref{line-c-AF}). We conclude that in Fig. \ref{fig4} the P phase
exists outside and AF phase inside lines (a) and (e), which meet
smoothly at the tricritical point TCP. The P phase becomes T unstable
below line (b), and AF phase below line (c). Notice that the lines (b)
and (c) are discontinuous across first-order transition
\cite{fyodorov87b}. It is likely that the first-order transition line
will change in this part of phase diagram once the
replica-symmetry-breaking solutions are considered for P and AF
phases.


\section{Conclusions}

In this paper we have investigated the stability of the RS symmetric
solution of the two-sublattice generalization of the SK infinite-range
spin-glass model.  We have derived stability conditions for
transversal fluctuations in agreement with previous investigations,
and we have extended previous study of the stability against
longitudinal or anomalous fluctuations. The eigenvalues associated
with such perturbations are in general complex. We generalized the
usual stability condition by requiring the real part of these
eigenvalues to be positive. The necessary and sufficient stability
conditions were found using the Hurwitz criterion for all the roots of
the secular equation to have positive real parts.  These conditions
allowed us to select one RS solution among those that are transversally
stable.  We believe that the generalized stability condition should
also be useful in other spin glass models where eigenvalues associated
with longitudinal and anomalous perturbations become complex.


\section*{\bf Acknowledgments}

The authors acknowledge partial financial support from the Brazilian
Government Agencies FINEP and CNPq.


\newpage
\newpage
\newpage
\newpage
\newpage


\newpage

\begin{thebibliography}{10}

\bibitem{sherrington75}
  D. Sherrington and S. Kirkpatrick, Phys. Rev. Lett. {\bf 35},  1792  (1975).

\bibitem{binder86}
  K. Binder and A.~P. Young, Rev. Mod. Phys. {\bf 58},  801  (1986).

\bibitem{mezard87}
  M. M{\'e}zard, G. Parisi, and M.~A. Virasoro, {\em Spin Glass Theory and
    Beyond} (World Scientific, Singapore, 1987).

\bibitem{fischer91}
  K.~H. Fischer and J.~H. Hertz, {\em Spin Glasses} (Cambridge University Press,
  Cambridge, 1991).

\bibitem{almeida78}
  J.~R.~L. de~Almeida and D.~J. Thouless, J. Phys. A {\bf 11},  983  (1978).

\bibitem{parisi79}
  G. Parisi, Phys. Lett. {\bf 73A},  203  (1979).

\bibitem{parisi80a}
  G. Parisi, J. Phys. A {\bf 13},  L115  (1980).

\bibitem{parisi80b}
  G. Parisi, J. Phys. A {\bf 13},  1101  (1980).

\bibitem{parisi80c}
  G. Parisi, J. Phys. A {\bf 13},  1887  (1980).

\bibitem{korenblit85}
  I.~Y. Korenblit and E.~F. Shender, Zh. Eksp. Teor. Fiz. {\bf 89},  1785
  (1985), [Sov. Phys. JETP {\bf 62}, 1030 (1985)].

\bibitem{fyodorov87a}
  Y.~V. Fyodorov, I.~Y. Korenblit, and E.~F. Shender, J. Phys. C {\bf 20},  1835
  (1987).

\bibitem{fyodorov87b}
  Y.~V. Fyodorov, I.~Y. Korenblit, and E.~F. Shender, Europhys. Lett. {\bf 4},
  827  (1987).

\bibitem{takayama88}
  H. Takayama, Prog. Theor. Phys. {\bf 80},  827  (1988).

\bibitem{bertrand82}
  D. Bertrand, A.~R. Fert, M.~C. Schmidt, F. Bensamka, and S. Legrand, J. Phys. C
  {\bf 15},  L883  (1982).

\bibitem{wong85a}
  P. zen Wong, S. von Molnar, T.~T.~M. Palstra, J.~A. Mydosh, H. Yoshizawa, S.~M.
  Shapiro, and A. Ito, Phys. Rev. Lett. {\bf 55},  2043  (1985).

\bibitem{wong85b}
  P. zen Wong, H. Yoshizawa, and S.~M. Shapiro, J. Appl. Phys. {\bf 57},  3462
  (1985).

\bibitem{yoshizawa87}
  H. Yoshizawa, S. Mitsuda, H. Aruga, and A. Ito, Phys. Rev. Lett. {\bf 59},
  2364  (1987).

\bibitem{yoshizawa89}
  H. Yoshizawa, S. Mitsuda, H. Aruga, and A. Ito, J. Phys. Soc. Jpn. {\bf 58},
  1416  (1989).

\bibitem{yoshizawa94}
  H. Yoshizawa, H. Mori, H. Kawano, H. Aruga-Katori, S. Mitsuda, and A. Ito, J.
  Phys. Soc. Jpn. {\bf 63},  3145  (1994).

\bibitem{dominicis85}
  C. de~Dominicis and I. Kondor,  in {\em Applications of Field Theory to
    Statistical Mechanics}, Vol.~216 of {\em Lecture Notes in Physics}, edited by
  L. Garrido (Springer-Verlag, Berlin, 1985), pp.\ 91--106.

\bibitem{lage82}
  E.~J.~S. Lage and J.~R.~L. de~Almeida, J. Phys. C {\bf 15},  L1187  (1982).

\bibitem{costa94a}
  F.~A. da~Costa, C.~S.~O. Yokoi, and S.~R.~A. Salinas, J. Phys. A {\bf 27},
  3365  (1994).

\bibitem{uspensky48}
  J.~V. Uspensky, {\em Theory of Equations} (McGraw-Hill Book Company, Inc., New
  York, 1948).

\bibitem{toulouse80}
  G. Toulouse, J. Phys. (Paris) Lett. {\bf 41},  L  (1980).

\end{thebibliography}
\end{document}